\documentclass[twoside,12pt]{article}
\usepackage{epsfig}

\newcommand{\be}{\begin{equation}}
\newcommand{\ee}{\end{equation}}
\newcommand{\bea}{\begin{eqnarray}}
\newcommand{\eea}{\end{eqnarray}}

\begin{document}

\title{ \vspace{1cm} Neutrinoless double beta decay studied with configuration mixing methods}
\author{Tom\'as R. Rodr\'iguez,$^{1,2,3}$ and Gabriel Martinez-Pinedo,$^{1}$\\
\\
$^1$GSI Helmholtzzentrum f\"ur Schwerionenforschung, D-64259 Darmstadt, Germany\\ 
$^2$Dpto. F\'isica Te\'orica, Universidad Aut\'onoma de Madrid,
E-28049 Madrid, Spain\\
$^3$CEA, Irfu, SPhN, Centre de Saclay, F-911191
  Gif-sur-Yvette, France}
\maketitle
\begin{abstract} We study neutrinoless double beta decay of several isotopes with state-of-the-art beyond self-consistent mean field methods to compute the nuclear matrix elements (NME). Generating coordinate method with particle number and angular momentum projection (GCM+PNAMP) is used for finding mother and granddaughter states and evaluating transition operators between different nuclei. We analyze explicitly the role of the deformation, pairing and configuration mixing in the evaluation of the NME. 
\end{abstract}
Double beta decay is a process where an even-even isotope decays into a nucleus with two less (more) neutrons (protons) with the emission of either two electrons and two neutrinos ($2\nu\beta\beta$) or only two electrons ($0\nu\beta\beta$). This process becomes the only decaying mode for those nuclei where the single beta decay to the odd-odd neighbor is energetically forbidden \cite{Avignone.Elliott.Engel:2008}. While the first mode has been already observed for several nuclei with half-lives of $\sim 10^{19-21}$~y, there is only one controversial claim of detection of neutrinoless double beta decay \cite{Klapdor-Kleingrothaus.Krivosheina.ea:2004}. This mode is very relevant because can only occur, beyond the standard model, if neutrinos are Majorana particles. Moreover, the inverse of the half-life of this process is directly related to the absolute mass of the neutrinos \cite{Avignone.Elliott.Engel:2008}:
\begin{equation}
\left[T^{0\nu\beta\beta}_{1/2}(0^{+}\rightarrow 0^{+})\right]^{-1}=G_{01}\left|M^{0\nu}\right|^{2}\left(\frac{\langle m_{\beta\beta}\rangle}{m_{e}}\right)^{2}
\label{halflife}
\end{equation}
where $\langle m_{\beta\beta}\rangle$ is the effective Majorana neutrino mass, $m_{e}$ is the electron mass, $G_{01}$ is a kinematical space factor and finally, $M^{0\nu}$ is the nuclear matrix element (NME). Due to the relevance of this process in particle and nuclear physics, there are several experiments devoted to search for $0\nu\beta\beta$ decay of different possible emitters \cite{Ejiri:2010}. From the theoretical point of view, the most important quantity to determine is the NME. Several nuclear structure methods have been applied to compute $0\nu\beta\beta$ NME being the Interacting Shell Model (ISM) \cite{Caurier.Menendez.ea:2008,Menendez.Poves.ea:2009} and proton-neutron Quasiparticle Random Phase Approximation (QRPA) in different versions \cite{Simkovic.Pantis.ea:1999,Simkovic.Faessler.ea:2008,Kortelainen.Suhonen:2007,Suhonen.Civitarese:2010} the most used ones. Additionally, calculations performed with two other methods - angular momentum Projected Hartree Fock Bogoliubov (PHFB) with a pairing plus quadrupole schematic interaction \cite{Chaturvedi.Chandra.ea:2008,chandra:2009} and Interacting Boson Model (IBM) \cite{Barea.Iachello:2009}- have been reported. Recently, we have proposed to study $M^{0\nu}$ matrix elements with energy density functional methods including beyond mean field effects \cite{arxiv}. In this contribution, we present the calculation of NME, total Gamow-Teller (GT) strengths $S_{+(-)}$ and Ikeda sum rule within this framework. We also show results for most of the isotopes that are considered as best candidates for detecting $0\nu\beta\beta$ decay giving a detailed analysis for $^{48}$Ca, $^{76}$Ge and $^{150}$Nd of interest for CANDLES \cite{CANDLES}, GERDA \cite{GERDA} and SNO+ \cite{SNO+} experiments respectively.

Energy density functional (EDF) methods have been extensively applied for describing properly many properties along the whole nuclear chart (see Ref. \cite{Bender.Heenen.Reinhard:2003} for a review). In this work we use an EDF based on the Gogny D1S interaction \cite{Berger.Girod.Gogny:1984} where particle number and rotational symmetries are restored and shape mixing (axial quadrupole degree of freedom) is taken into account within the framework of the Generating Coordinate Method (GCM+PNAMP functional). Contrary to other methods, single particle energies and residual interactions come from the same functional self-consistently. We assume the closure approximation for calculating NME's because at present it is not possible to compute the intermediate odd-odd nucleus within this framework. Then, $M^{0\nu}$ of Eq. \ref{halflife} can be evaluated as \cite{Avignone.Elliott.Engel:2008}:
\begin{equation}
M^{0\nu}=\langle 0^{+}_{f}|\hat{M}^{0\nu}|0^{+}_{i}\rangle\,\,;\hat{M}^{0\nu}=-\left(\frac{g_{V}}{g_{A}}\right)^{2}\hat{M}^{0\nu}_{F}+\hat{M}^{0\nu}_{GT}-\hat{M}^{0\nu}_{T}
\end{equation}
where $|0^{+}_{i/f}\rangle$ are the initial and final ground states, $g_{V}=1, g_{A}=1.25$ are the vector and axial vector coupling constants and $\hat{M}_{F/GT/T}$ are the Fermi, Gamow-Teller and Tensor two-body operators. In this work, tensor term is neglected~\cite{Menendez.Poves.ea:2009,Kortelainen.Suhonen:2007} and F and GT operators can be written as:
\begin{equation}
  \hat{M}^{0\nu}_{F}=\hat{V}_{F}\hat{t}^{(1)}_{-}\hat{t}^{(2)}_{-},\,\,
  \hat{M}^{0\nu}_{GT}=\hat{V}_{GT} (\hat{\vec{\sigma}}^{(1)} \cdot
  \hat{\vec{\sigma}}^{(2)})\hat{t}^{(1)}_{-}\hat{t}^{(2)}_{-}   
  \label{NME}
\end{equation} 
Here $\hat{t}_{-}$ is the isospin ladder operator that changes
neutrons to protons, $\hat{\vec{\sigma}}$ are the Pauli matrices acting
on the spin part and $\langle\vec{r}_{1}\vec{r}_{2} |\hat{V}_{F/GT}|
\vec{r}_{1}\vec{r}_{2} \rangle=v_{F/GT}(|\vec{r}_{1}-\vec{r}_{2}|)$
are local potentials that depend on the relative coordinate of the
nucleons involved in the decay,
$r\equiv|\vec{r}_{1}-\vec{r}_{2}|$. The functions $v_{F/GT}(r)$ are
the so-called neutrino potentials including high order
currents~\cite{Simkovic.Pantis.ea:1999}, nucleon finite size
corrections~\cite{Simkovic.Pantis.ea:1999} and radial short range correlations treated
within the Unitary Correlator Method~\cite{Feldmeier.Neff.ea:1998,Kortelainen.Suhonen:2007} (see
Ref.~\cite{Menendez.Poves.ea:2009} for further details). 
We now describe the method for calculating the ground states $|0^{+}_{i/f}\rangle$ and observables using these wave functions. For the sake of simplicity, we formulate the theoretical framework in terms of operators and wave functions although a derivation in terms of densities can be found in Refs.~\cite{Rodriguez.Egido.Robledo:2002,lacroix.duguet.bender:2009}. The starting point is the GCM wave function:
\begin{equation}
  |0^{+}\rangle=\sum_{\beta}g_{\beta}P^{I=0}P^{N}P^{Z}|\Phi_{\beta}\rangle
\end{equation}
where $P^{I=0}=\int_{0}^{\pi/2} e^{-i\hat{J_{y}}b}\sin b\,db$, $P^{N}=\frac{1}{2\pi}\int^{2\pi}_{0}e^{i\varphi(\hat{N}-N)}d\varphi,P^{Z}=\frac{1}{2\pi}\int^{2\pi}_{0}e^{i\varphi(\hat{Z}-Z)}d\varphi$ are the corresponding angular momentum 
($I=0$) -for axial symmetric wave functions- and particle number projectors~\cite{Ring.Schuck:1980}. The
intrinsic axial symmetric Hartree-Fock-Bogoliubov (HFB) wave functions
$|\Phi_{\beta}\rangle$ are solutions to the variation after particle
number projection equations constrained to a given value of the quadrupole deformation,
$\beta$~\cite{Anguiano.Egido.Robledo:2001,Rodriguez.Egido:2007}. These intrinsic wave functions are vacuum for the quasiparticle operators $\hat{\alpha}_{\gamma}$ that are defined by the HFB transformation~\cite{Ring.Schuck:1980}:
\begin{equation}
\hat{\alpha}^{\dagger}_{\gamma}=\sum_{\delta}U_{\delta\gamma}\hat{c}^{\dagger}_{\delta}+V_{\delta\gamma}\hat{c}_{\delta}
\end{equation}
with $U,V$ being the HFB variational coefficients and $\hat{c}^{\dagger}_{\delta} (\hat{c}_{\delta})$ creation (annihilation) operators in the arbitrary single particle basis in which we expand the many body intrinsic wave functions. In our case, this basis corresponds to a spherical harmonic oscillator potential solved in cartesian coordinates that includes eleven major oscillator shells.
Now, the coefficients $g_{\beta}$ are found by solving
the Hill-Wheeler-Griffin (HWG) equation
\cite{Ring.Schuck:1980,Rodriguez.Egido.Robledo:2002,lacroix.duguet.bender:2009}. First, for each nucleus we transform the
non-orthogonal set of wave functions
$\left\{P^{I=0}P^{N}P^{Z}|\Phi_{\beta}\rangle\right\}$ into an
orthonormal one
$\left\{|\Lambda\rangle=\sum_{\beta}\frac{u_{\Lambda,\beta}}{\sqrt{n_{\Lambda}}}P^{I=0}P^{N}P^{Z}|\Phi_{\beta}\rangle\right\}$
by diagonalizing the norm overlap matrix,
$\sum_{\beta'}\langle\Phi_{\beta}|P^{I=0}P^{N}P^{Z}|\Phi_{\beta'}\rangle
u_{\Lambda,\beta'}=n_{\Lambda}u_{\Lambda,\beta}$.  In this basis, the
HWG equation reads: $\sum_{\Lambda'}\varepsilon_{\Lambda\Lambda'}
G^{a}_{\Lambda'}=E^{a}G^{a}_{\Lambda}$, where
$\varepsilon_{\Lambda\Lambda'}$ are the so-called energy
kernel~\cite{Rodriguez.Egido:2007,Rodriguez.Egido.Robledo:2002,lacroix.duguet.bender:2009}.  Finally, the coefficients for the
lowest eigenvalue are used to compute both the so-called collective
wave functions $F(\beta)=\sum_{\Lambda}
G^{0}_{\Lambda}u_{\Lambda,\beta}$ - probability distribution for the
state to have a given deformation - and all observables. For scalar -under rotation- operators $\hat{O}$, matrix elements between $|0^{+}_{i/f}\rangle$ states can be calculated with the expression:  
\begin{equation}
    O=\sum_{\Lambda_{i}\Lambda_{f}}\sum_{\beta_{i}\beta_{f}}
    \left(\frac{u^{*}_{\Lambda_{f},\beta_{f}}}{\sqrt{n_{\Lambda_{f}}}}\right) 
    G^{0*}_{\Lambda_{f}} \langle \Phi_{\beta_{f}}| P^{N_{f}}
    P^{Z_{f}} \hat{O}P^{I=0}P^{N_{i}}P^{Z_{i}}|
    \Phi_{\beta_{i}} \rangle G^{0}_{\Lambda_{i}} 
   \left(\frac{u_{\Lambda_{i},\beta_{i}}}{\sqrt{n_{\Lambda_{i}}}}\right). 
    \label{observables}
  \end{equation}
In this work we will focus our interest on calculating $0\nu\beta\beta$ NME's and also ground state energies, charge radii and total Gamow-Teller strengths $S_{+/-}$. We will compare these quantities with the available experimental data to check the reliability of the method. 
In particular, the dependence on deformation of the $0\nu\beta\beta$ NME's is evaluated with the kernel of Eq. \ref{observables} for the specific case of $\hat{O}=\hat{M}^{0\nu}_{\xi}$ ($\xi=F/GT$):
\begin{equation}
M^{0\nu}_{\xi}(\beta_{i},\beta_{f})=\frac{\langle \Phi_{\beta_{f}}| P^{N_{f}}
    P^{Z_{f}} \hat{M}^{0\nu}_{\xi}P^{I=0}P^{N_{i}}P^{Z_{i}}|
    \Phi_{\beta_{i}} \rangle}{\sqrt{\langle \Phi_{\beta_{f}}|P^{I=0}P^{N_{f}}P^{Z_{f}}|
    \Phi_{\beta_{f}} \rangle}\sqrt{\langle \Phi_{\beta_{i}}|P^{I=0}P^{N_{i}}P^{Z_{i}}|
    \Phi_{\beta_{i}} \rangle}}
\label{m_Gt}    
\end{equation}
Finally, we evaluate the total GT strength calculating the matrix elements of the operator $\hat{\vec{O}}^{\pm}_{GT}=\hat{\vec{\sigma}}\hat{t}_{+/-}$ between the ground state $|0^{+}\rangle$ of the even-even initial nucleus and the $1^{+}_{m}$ states in the odd-odd final nucleus:
\begin{equation}
S_{+/-}=\sum_{1^{+}_{m}}|\langle 1^{+}_{m}|\hat{\vec{O}}^{\pm}_{GT}|0^{+}\rangle|^{2}=\langle0^{+}|(\hat{\vec{O}}^{\pm}_{GT})^{\dagger}\cdot\hat{\vec{O}}^{\pm}_{GT}|0^{+}\rangle
\label{gt}
\end{equation} 
In the r.h.s of Eq. \ref{gt} we have taken into account the completeness relation of the final states -$\sum_{1^{+}_{m}}|1^{+}_{m}\rangle\langle1^{+}_{m}|=1$- that leads to a two-body form of the corresponding operators:
\begin{equation}
\hat{S}_{+}=3\hat{Z}-\hat{O}_{\rm{2b}}\,;\,\hat{S}_{-}=3\hat{N}-\hat{O}_{\rm{2b}}\,;\,\hat{O}_{\rm{2b}}=\sum_{\alpha\zeta\gamma\delta}\left(\vec{\sigma}\right)_{\alpha\zeta}\left(\vec{\sigma}\right)_{\gamma\delta}\hat{b}^{\dagger}_{\alpha}\hat{b}_{\delta}\hat{a}^{\dagger}_{\gamma}\hat{a}_{\zeta}
\label{smasmenos}
\end{equation}
where $\left(\vec{\sigma}\right)_{\alpha\zeta}$ are the matrix elements of the Pauli operators in the harmonic oscillator single particle basis and  $\hat{b}_{\alpha}$ ($\hat{a}^{\dagger}_{\zeta}$) annihilates (creates) a neutron (proton) in the orbit $\alpha$~($\zeta$). From the above expression we can also see that Ikeda's sum rule ($S_{-}-S_{+}=3(N-Z)$) is fulfilled if particle number conserving wave functions are considered. 

\begin{figure}[h!]
\centering
\includegraphics[width=1.0\columnwidth]{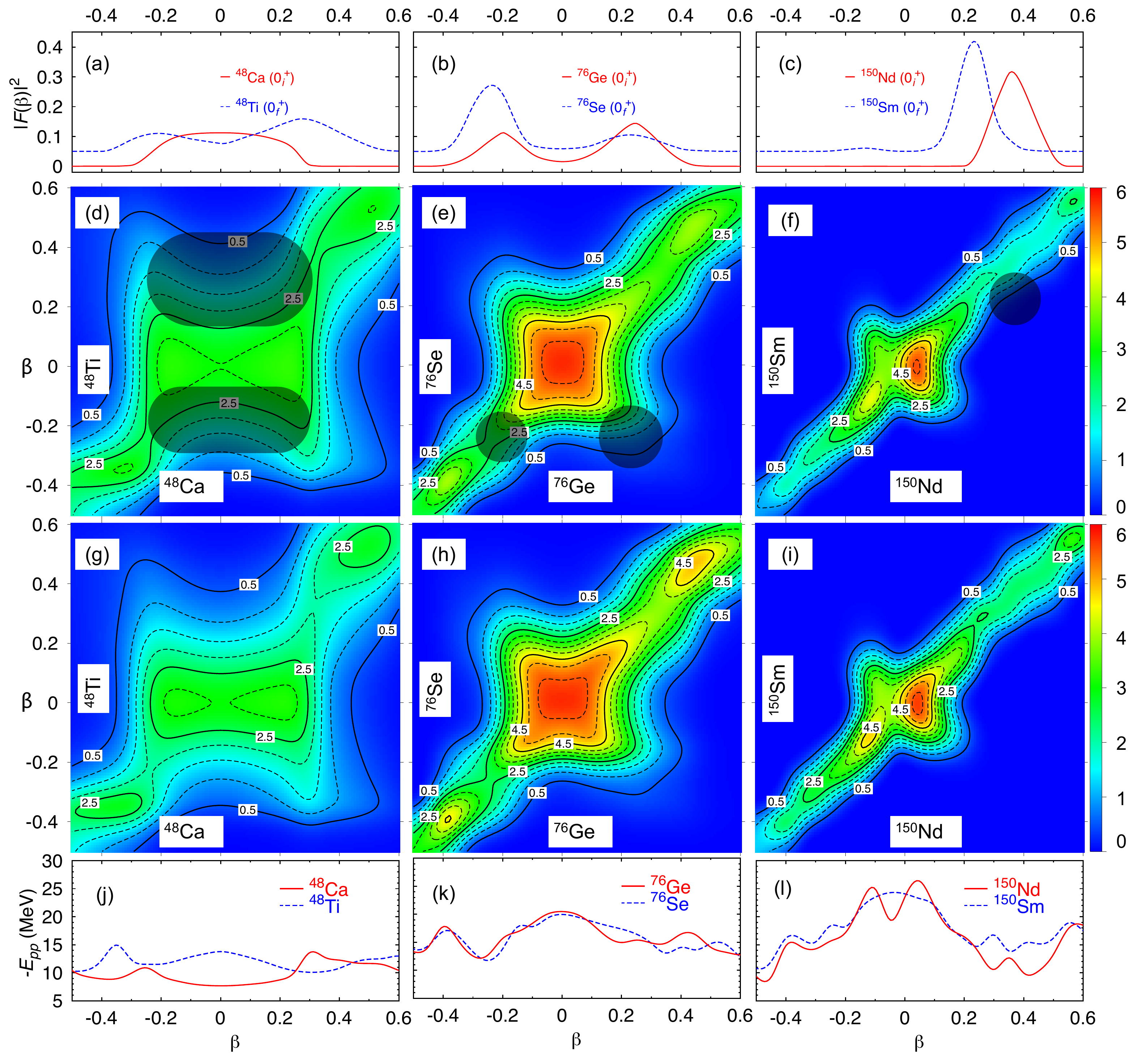}
\caption{(a)-(c) Collective wave functions, GT intensity with, (d)-(f) full and, (g)-(i) constant spatial dependence and (j)-(l) pairing energies for (left) $A=48$, (middle) $A=76$ and (right) $A=150$ decays. Shaded areas corresponds to regions explored by the collective wave functions.}
\label{Figure1}
\end{figure}

Before evaluating the $0\nu\beta\beta$ NME's within the framework described above, we check the reliability of the method comparing theoretical and experimental results for masses, charge radii and total GT strengths. We observe a very nice agreement between the computed values and the experimental ones. For example, relative errors for masses and radii are less than 1.5$\%$ in all cases. It is important to note that total GT strengths $S_{+/-}$ have been quenched by a factor $(0.74)^{2}$ as it is usually done in ISM \cite{SM_RMP} and QRPA \cite{Alvarez-Rodriguez.Sarriguren.ea:2004} calculations.  
\begin{table}[h]
   \centering
   \begin{small}
   \begin{tabular}{ccccccc} 
   \hline \hline
Isotope & $BE^{th}$ (MeV) & $BE^{exp}$ (MeV) \cite{audi}& $R^{th}$ (fm) & $R^{exp}$ (fm) \cite{radii}& $S^{theo}_{-/+}$ & $S^{exp}_{-/+}$ \\
\hline
$^{48}$Ca & 420.623 & 415.991  & 3.465 & 3.473 & 13.55 & (14.4$\pm$2.2 \cite{CaCER_09})\\
$^{48}$Ti & 423.597 & 418.699  & 3.557 & 3.591 & 1.99 & (1.9 $\pm$ 0.5 \cite{CaCER_09})\\
$^{76}$Ge & 664.204 & 661.598  & 4.024 & 4.081 & 20.97 & (19.89 \cite{Anderson_89})\\
$^{76}$Se & 664.949 & 662.072  & 4.074 & 4.139 & 1.49 & (1.45 $\pm$ 0.07 \cite{SeCER_97})\\
$^{82}$Se & 716.794 & 712.842  & 4.100 & 4.139 & 23.56 & (21.91 \cite{Anderson_89})\\    
$^{82}$Kr & 717.859 & 714.273  & 4.130 & 4.192 & 1.24 & \\
$^{96}$Zr & 829.432 & 828.995  & 4.298 & 4.349 & 27.63 & \\  
$^{96}$Mo & 833.793 & 830.778  & 4.319 & 4.384 & 2.56 & (0.29 $\pm$ 0.08 \cite{MoCER_08})\\
$^{100}$Mo & 861.526 & 860.457 & 4.372 & 4.445 & 27.87 & (26.69 \cite{Anderson_89})\\
$^{100}$Ru & 864.875 & 861.927 & 4.388 & 4.453 & 2.48 & \\
$^{116}$Cd & 988.469 & 987.440 & 4.556 & 4.628 & 34.30 & (32.70 \cite{Anderson_89})\\
$^{116}$Sn & 991.079 & 988.684 & 4.567 & 4.626 & 2.61 & (1.09$^{+0.13}_{-0.57}$ \cite{SnCER_05})\\
$^{124}$Sn & 1051.668 & 1049.96 & 4.622 & 4.675 & 40.65 & \\
$^{124}$Te & 1051.562 & 1050.69 & 4.664 & 4.717 & 1.63 & \\
$^{128}$Te & 1082.257 & 1081.44 & 4.686 & 4.735 & 40.48 &(40.08 \cite{Anderson_89})\\
$^{128}$Xe & 1080.996 & 1080.74 & 4.723 & 4.775 & 1.45 & \\
$^{130}$Te & 1096.627 & 1095.94 & 4.695 & 4.742 & 43.57 & (45.90 \cite{Anderson_89})\\
$^{130}$Xe & 1097.245 & 1096.91 & 4.732 & 4.783 & 1.19 & \\
$^{136}$Xe & 1143.333 & 1141.88 & 4.756 & 4.799 & 46.71 & \\
$^{136}$Ba & 1143.202 & 1142.77 & 4.786 & 4.832 & 0.96 & \\
$^{150}$Nd & 1234.512 & 1237.45 & 5.034 & 5.041 & 50.32 &\\
$^{150}$Sm & 1235.936 & 1239.25 & 5.041 & 5.040 & 1.45 & \\
\hline
   \end{tabular}
   \end{small}
   \caption{Masses, rms charge radii and total Gamow-Teller strengths $S_{-(+)}$ for mother (granddaughter) calculated with Gogny D1S GCM+PNAMP functional compared to experimental values. Theoretical values for $S_{+/-}$ are quenched by a factor (0.74)$^{2}$.}
   \label{table1}
\end{table}
We now study the dependence of the $0\nu\beta\beta$ NME's on deformation and pairing correlations of the isotopes involved in the process. In Figure \ref{Figure1}(d)-(f) we plot the magnitude of the GT contribution as a function of the quadrupole deformation of mother and granddaughter nuclei for $A=48$, 76 and 150 decays (Eq. \ref{m_Gt}) -Fermi contributions have a similar form and are not shown. We observe that the intensity is distributed rather symmetrically along the $\beta_{i}=\beta_{f}$ direction having its larger values in regions close to the spherical points. This region is wider for the $A=48$ case than for $A=76$ and 150. Apart from this area, the NME is strongly suppressed whenever the difference in deformation of the two nuclei is large. This has been also noticed in ISM \cite{Menendez.Poves.ea:2008} and PHFB calculations \cite{Chaturvedi.Chandra.ea:2008,chandra:2009} for $0\nu\beta\beta$ and QRPA calculations \cite{Alvarez-Rodriguez.Sarriguren.ea:2004} for $2\nu\beta\beta$. Nevertheless, we also obtain maxima and minima along the diagonal $\beta_{i}=\beta_{f}$ finding a non-trivial structure. Concerning the magnitude of the GT NME's, we obtain less intensity in $A=48$ than in the rest, having maximum values of 3.3, 5.7 and 5.8 for $A=48$, 76 and 150 respectively. In order to shed light on this structure we represent in Fig. \ref{Figure1}(g)-(i) the results for a modified GT operator where all the involved spatial dependence of the neutrino potentials are substituted by constant potentials -$V_{GT}=V_{GT,0}=-1.5,-2.3,-2.2$ for $A=48,76,150$ respectively. With this assumption, GT $0\nu\beta\beta$ matrix elements reduce to the $2\nu\beta\beta$ ones evaluated in the closure approximation because $\hat{M}^{2\nu}_{\rm{cl}}\propto (\hat{\vec{\sigma}}^{(1)} \cdot\hat{\vec{\sigma}}^{(2)})\hat{t}^{(1)}_{-}\hat{t}^{(2)}_{-} $ (see Eq. \ref{NME}). We observe that the structure as a function of the deformation is reproduced with this simplified operator indicating that $M_{\rm{GT}}^{0\nu}(\beta_{i},\beta_{f})$ matrix elements are almost proportional to $M_{\rm{cl}}^{2\nu}(\beta_{i},\beta_{f})$. In addition, we can also relate the structure of the operator given in Fig. \ref{Figure1}(d)-(f) with the amount of pairing correlations in the mother and granddaughter nuclei separately (pairing energy $-E_{pp}$ \cite{Ring.Schuck:1980}) represented in Fig. \ref{Figure1}(j)-(l). We can observe a direct correlation between maxima and minima found both in the pairing energy and in the intensity of the GT contribution. Furthermore, the value of the strength is bigger with larger pairing correlations, in agreement with ISM calculations where zero seniority calculations give the largest values of the NME's \cite{Caurier.Menendez.ea:2008}. 

\begin{figure}[h!]
\centering
\includegraphics[width=0.6\columnwidth]{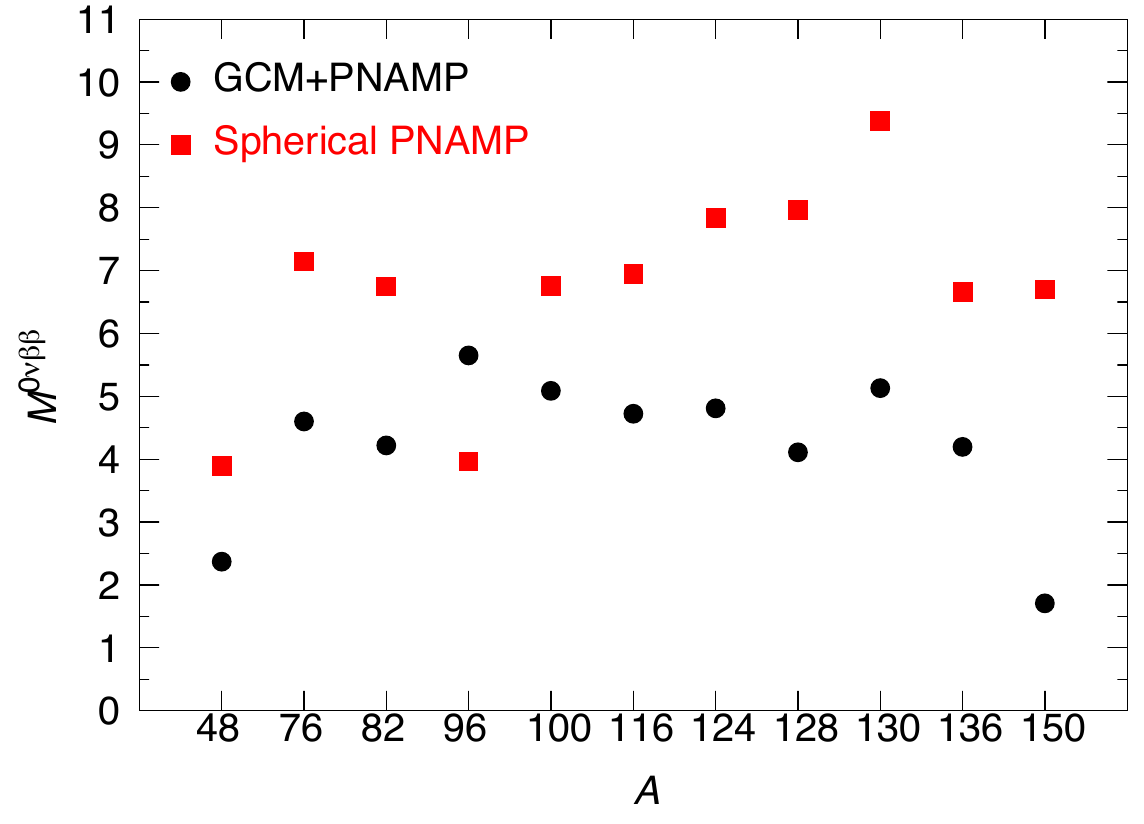}
\caption{$0\nu\beta\beta$ matrix elements evaluated with full configuration mixing (circles) and assuming spherical symmetry (squares).}
\label{Figure2}
\end{figure}
Finally, we have to take into account the effect of configuration mixing in the calculations. In Fig. \ref{Figure1}(a)-(c) we show the collective wave functions for initial and final states of $A=48,76,150$ decays respectively. Here, we observe for $^{48}$Ca a rather constant distribution of probability between $\beta\approx[-0.2,+0.2]$ with the maximum at the spherical point; for $^{48}$Ti we find two maxima in $\beta\approx-0.2$ and $\beta\approx+0.3$ and a minimum in $\beta=0$ giving in average a very slightly prolate deformed state; this is also found for $^{76}$Ge with maxima at $\beta\approx\pm0.20$ while $^{76}$Se is mostly oblate deformed with a peak at $\beta\approx-0.25$. Finally, both $^{150}$Nd and $^{150}$Sm are well prolate deformed but their wave functions peak at different deformations ($\beta\approx+0.40$ and $\beta\approx+0.25$, respectively). According to Eq. \ref{observables}, the final results depend on the convolution of the collective wave functions with the $0\nu\beta\beta$ matrix elements as a function of deformation. In Fig. \ref{Figure1}(d)-(f) we show schematically -shaded circles- the areas of the GT intensity explored by the collective wave functions. We observe, on the one hand, that configuration mixing is very important in the final result because several shapes can contribute to the value of NME, especially in $A=48$ and 76. On the other hand, we see that the regions with largest values of the GT intensity are excluded by the collective wave functions. For example, calculations assuming spherical symmetry give systematically larger NME -except for $A=96$- as we show in Figure \ref{Figure2}. 

To summarize, we have presented a method for calculating $0\nu\beta\beta$ nuclear matrix elements based on Gogny D1S Energy Density Functional including beyond mean field effects such as symmetry restoration and shape mixing. We have validated our method comparing theoretical and experimental ground state properties. Then, we have studied in detail the dependence on deformation of NME in $A=48,76,150$ decays relating the structure obtained with the pairing correlations in the mother and granddaughter nuclei. We have also found a close connection between the strength of the NME's and the results obtained assuming a constant dependence in the spatial part of the transition operators. Finally, we have pointed out the relevance of having configuration mixing in the calculations.  

We thank A. Poves, J. Men\'{e}ndez, J.L. Egido, K. Langanke, T. Duguet and F. Nowacki for fruitful discussions. TRR is supported by the Programa de Ayudas para Estancias de Movilidad Posdoctoral 2008 and FPA2009-13377-C02-01 (MICINN). GMP is partly supported by the DFG through contract SFB 634, by the ExtreMe Matter Institute EMMI and by the Helmholtz International Center for FAIR.

\end{document}